\documentclass[12pt]{article}
\usepackage{amssymb}
\usepackage{epsfig}
\usepackage{graphicx}
\usepackage{amsmath}
\usepackage{verbatim}

\csname @addtoreset\endcsname{equation}{section}

\begin{document}

\begin{titlepage}

\title{\bf\Large $E2$ Instanton Effects and Higgs Physics
 In Intersecting Brane Models \vspace{18pt}}
\author{\normalsize  Mingxing Luo and Sibo Zheng\vspace{12pt}\\
{\it\small Zhejiang Institute of Modern Physics, Department of Physics,}\\
{\it\small Zhejiang University, Hangzhou 310027, P.R. China}\\
{\small e-mail: { ~\it luo@zimp.zju.edu.cn,
zhengsibo@zimp.zju.edu.cn}}}

\date{}
\maketitle  \voffset -.2in \vskip 2cm \centerline{\bf Abstract}
\vskip .4cm String instanton effects in Higgs physics are
discussed through a type IIA model based on $T^{6}/(Z^{2}\times
Z^{'2})$ orentifold compactifaction. By inclusion of rigid
$E2$-branes, the model exhibits an MSSM-like spectrum, as well as
extra $\mu$ and quartic Higgs couplings. These extra couplings are
induced  via $E2$ instantons non-perturbatively. Setting the
string scale at $10^{18}$ GeV, one gets interesting TeV Higgs
physics. In particlular, the tree-level Higgs mass can be uplifted
substantially.

\end{titlepage}
\newpage
\setcounter{page}{1}

\section{Introduction}

Recently, string instanton effects have been intensively
explored in moduli stabilization of flux induced compactifications
\cite{BKLS,BCLS,Grana} (and references therein) and string
phenomenology, especially for non-perturbative generation of
right-handed neutrino masses and $\mu$ term in intersecting
brane models \cite{Buican,CRW,CL,IU}. These
non-perturbative effects come from nonzero global charges
$Q_{a}=N_{a}\Xi\circ (\Pi_{a}-\Pi^{'}_{a})$ carried by the
instanons, which lead to interesting charged matter couplings \cite{BCW}
(for recent reviews, see for example, \cite{ ABLS,CRW08}).
Setting the string scale at the order of $10^{18}$ GeV,
one finds $m_{\nu}$ and $\mu$ in acceptable ranges without any fine-tuning.

In addition to the $\mu$ term, there is another important coupling
in Higgs physics, the quartic coupling, which controls  Higgs boson
masses. In the minimal supersymmetric standard model (MSSM), the
tree-level mass of the lighest Higgs particle $h$ is well below
the LEPII bound. To make ends meet, one needs substantial
radiative contribution to $m_{h}$ which is dominated by the stop
quark \cite{oneloop,Quiros}. In order to obtain the desired
up-lifting, both the stop mass and the mixing have to be large.
And this greatly constrains the parameter space in the MSSM and
aggraviates fine tuning problems associated with soft mass terms.
This provides motivations to make extensions beyond the MSSM, such
as the next leading-order minimal supersymmetric standard model
(NMSSM) \cite{NMSSM} and beyond minimal supersymmetric standard
model (BMSSM)\cite{DTS,BCEN}. In certain examples, extra quaritc
Higgs couplings are present which modify tree-level Higgs masses.
Their significance is controlled by the mechanism of supersymmetry
breaking in hidden sector and the value of the associated mass
scale.

Motivated by the rich phenomenologies generated by stringy
instantons, in this paper we will discuss their effects on two important mass
scales in Higgs physics, i.e, the $\mu$ term and the mass scale $M$
associated with the quartic couplings, in $T^{6}/(Z^{2}\times
Z^{'2})$ orentifold compactifaction of type IIA theories \cite{setup}.
They are induced non-perturbatively via $E2$ instantons.
Setting the string GUT scale at $10^{18}$ GeV,
one gets interesting TeV Higgs physics.
In particlular, the tree-level Higgs mass can be uplifted substantially.

In section 2, a $\mathcal{N}=1$
supersymmetric model is constructed that exhibits an MSSM-like spectrum (including
the right-handed neutrino) with suitable wrapping numbers of $D6$ and
$E2$ branes. In section 3, we discuss the generations of $E2$-branes
induced $\mu$ term and quartic couplings. The structure of these
quartic terms are explicitly calculated. They obviously modify the
Higgs masses, which are expressed as expansions of a small parameter
$\varepsilon\sim \mu/M$, as shown in section 4. We conclude in
section 5.

\section{The setup}
We discuss an intersecting $D6$-branes model in $T^{6}/(Z^{2}\times
Z^{'2})$ orentifold of type IIA theories.
All the moduli are stabilized if non-perturbative $E2$-brane instanton
effects are taken into account \cite{BKLS,BCLS,Grana},
and standard model spectrum can be obtained by properly arranging the intersecting branes.
Shown in table 1 are the wrapping numbers of four-stack branes $a,b,c,d$.
The model carries gauge groups $U(3)_{a}\times
U(2)_{b}\times U(1)_{c}\times U(1)_{d}$, of which all the
$U(1)_{i}$ become massive by the Green-Schwarz mechanism except $U(1)_{Y}$,
\begin{eqnarray}{\label{hypercharge}}
Q_{Y}=\frac{1}{6}Q_{a}-\frac{1}{2}Q_{c}-\frac{1}{2}Q_{d}
\end{eqnarray}
The gauge groups then conforms to that of MSSM-like theories. The
intersecting number $I_{cd}=-3$ implies neutrinos $\nu_{R}$ are
also encoded. Shown in table 2 are the chiral spectra of theories
corresponding to wraping numbers in table 1.

For the model to be supersymmetric, each stack of branes
has to satisfy two conditions \cite{Shiu},
\begin{eqnarray}{\label{susy1}}
m_{x}^{1}m_{x}^{2}m_{x}^{3}-\sum_{I\neq J\neq
K}\frac{n_{x}^{I}n_{x}^{J}m_{x}^{K}}{U^{I}U^{J}}=0
\end{eqnarray}
and
\begin{eqnarray}{\label{susy2}}
n_{x}^{1}n_{x}^{2}n_{x}^{3}-\sum_{I\neq J\neq
K}m_{x}^{I}m_{x}^{J}n_{x}^{K}U^{I}U^{J}> 0
\end{eqnarray}
where $U^{I}=R_{Y}^{I}/R_{X}^{I}$ is the complex structure
modulus of $I$th torus with radii $R_{X}^{I},R_{Y}^{I}$.

Note that in table 1,
$N_{h}$ $D6$-branes and $N_{O}$ $O6$ branes are added to cancel the tadpoles,
\begin{eqnarray}{\label{tadpole}}
\sum_{a=1}^{K}N_{a}(\Pi_{a}+\Pi^{'}_{a})=N_{O}\Pi_{O6}
\end{eqnarray}
Also, stacks $a$ and $d$ are parallel in the transverse
directions. The open string modes stretching between them are
massive, of the order $L/(\sqrt{2\pi\alpha_{s}})$ ($L$ is the transverse
distance). So matter contents in table 2 are exact in the
effective theory below the string scale. In addition, two $E2$-branes
$M,N$ are embedded. We will see in the next section that they
non-perturbatively induce interesting small $\mu$ term and quartic
terms in Higgs physics, respectively.

\begin{table}
\begin{center}
\begin{tabular}{|c|c|c|c|}
 \hline
  $N_{i}$ & $(n^{1}_{i},m^{1}_{i})$ & $(n^{2}_{i},m^{2}_{i})$ & $(n^{3}_{i},m^{3}_{i})$ \\
  \hline
  $N_{a}=6$ & $(1,0)$ & $(3,1)$ & $(3,-1/2)$ \\
  $N_{b}=4$ & $(1,1)$ & $(1,0)$ & $(1,-1/2)$ \\
  $N_{c}=2$ & $(0,1)$ & $(0,-1)$ & $(2,0)$ \\
  $N_{d}=2$ & $(1,0)$ & $(3,1)$ & $(3,-1/2)$ \\
  $N_{h}=4$ & $(-2,1)$ & $(-3,1)$ & $(-3,1/2)$ \\
  $N_{O}=6$ & $(1,0)$ & $(1,0)$ & $(1,0)$ \\
  $E2_{M}$ & $(1,0)$ & $(1,-1)$ & $(1,1/2)$ \\
  $E2_{N}$ & $(n^{1}_{N},-n^{1}_{N})$ & $(n^{2}_{N},\frac{12n_{N}^{2}}{1-s})$ & $(\frac{6n_{N}^{1}(n_{N}^{2})^{3}}{1-s},
  \frac{1}{n^{1}_{N}n^{2}_{N}})$ \\
  \hline
\end{tabular}
\caption{ Wrapping numbers of $D6$-branes and $E2$-instantons which wrap on a rigid three-cycle on $Z^{2}\times
Z^{'2}$ toroidal orentifold.
$n^{1}_{N},n^{2}_{N}$ are real numbers ($
s=(n_{N}^{1})^{2}(n_{N}^{2})^{4}$). The model is supersymmetric
if $U_{3}=2U_{1}=-2U_{2}=1$. }
\end{center}
\end{table}

\begin{table}
\begin{center}
\begin{tabular}{|c|c|c|}
  \hline
  intersection numbers & matter & Rep \\
  \hline
  $I_{ab}=I_{ab^{*}}=3$ & $Q_{L}$ & $3(3,2)$ \\
  $I_{ac}=-3$ & $U_{R}$ & $3(\bar{3},1)$ \\
  $I_{ac^{*}}=3$ & $D_{R}$ & $3(\bar{3},1)$ \\
  $I_{db}=I_{db^{*}}=3$ & $L$ & $3(1,2)$ \\
  $I_{cd}=-3$ & $\nu_{R}$ & $3(1,1)$ \\
  $I_{cd}=3$ & $E_{R}$ & $3(1,1)$ \\
  $I_{bc}=-1$ & $H_{u}$ & $1(1,2)$ \\
  $I_{bc^{*}}=-1$ & $H_{v}$ & $1(1,2)$ \\
  \hline
\end{tabular}
\caption{Chiral matters spectrum for the wraping numbers in table
1.}
\end{center}
\end{table}

\section{Non-perturbative Higgs physics from $E2$ instanton }

To yield a non-perturbative $\mu$ term,
one assigns the following intersection numbers between $E2$-brane and
$D6_{b,c}$-branes
\begin{eqnarray}{\label{E2M1}}
I_{Mb}=-1,~~~I_{Mb^{*}}=0,~~~ I_{Mc}=I_{Mc^{*}}=1~~~ (I_{bc}<0)
\end{eqnarray}
The intersection number $I_{M\alpha}$ also has to satisfy,
\begin{eqnarray}{\label{E2M2}}
I_{M\alpha}-I_{M\alpha^{*}}=0,~~~(\alpha=a,d)
\end{eqnarray}
in order to exclude the extra charged zero modes. The wrapping numbers on
$E2_{M}$ are $2(1,0)(1,-1)(1,1/2)$, which are determined by the
constraints Eq. \eqref{E2M1} and Eq. \eqref{E2M2}, as shown in table 1.
The number of triangles on each torus is 1, contributing to
$H^{ij}_{u}\lambda_{a}^{i}\bar{\lambda}_{b}^{j}e^{-\mathcal{A}_{i}}$
and
$H^{ij}_{d}\lambda_{a}^{i}\bar{\lambda}_{b}^{j}e^{-\mathcal{A}_{i}}$
terms respectively for intersecting $(b,c)$ and $(b,c^{*})$ branes.
This generates a $\mu H_{u}H_{d}$ term non-perturbatively
in four-dimensional effective theory, as desired \cite{Buican,CL,IU}.

We now discuss the quartic operator $\frac{\lambda}{M}(H_{u}H_{d})^{2}$
and its implication for Higgs physics.
These operators were constructed in certain BMSSM examples. They can
greatly uplift Higgs masses when $M$ is in the range of $1\sim 10$ TeV.
Similar to the stringy instanton induced $\mu$ term as shown above,
it is possible to construct these quartic terms non-perturbatively.
That is, the roles played by hidden sectors to generate these operators in other models
can be totally replaced by stringy instanton effect in our model.

In order to exclude extra zero modes on $D6_{a,d}$-branes,
one has the constraints on the intersection number $E2_{N}$ and $D6$-branes
\begin{eqnarray}{\label{constraints}}
I_{N\alpha}&=&I_{N\alpha^{*}},~~~(\alpha=a,d)
\end{eqnarray}
and
\begin{eqnarray}{\label{E2N}}
I_{Nb}&=&-4,~~~I_{Nb^{*}}=0,~~~ I_{Nc}=I_{Nc^{*}}=2~~~ (I_{bc}<0)
\end{eqnarray}
which can be obtained by counting the numbers of charged zero modes that arise from strings strechting bewteen
the $E2_{N}$ and $D6_{b,c}$-branes.

As shown in table 2, the wrapping numbers of $E2_{N}$-brane are reprensented by two
integer $(n^{1}_{N},n^{2}_{N})$.
$E2_{N}$ also preserve the same supersymmetry as $D6$-branes, i.e,
the wrapping numbers of $E2_{N}$ satisfy the constraints
Eq. \eqref{susy1} and Eq. \eqref{susy2}.

\begin{figure}
\includegraphics[width=0.8\linewidth]{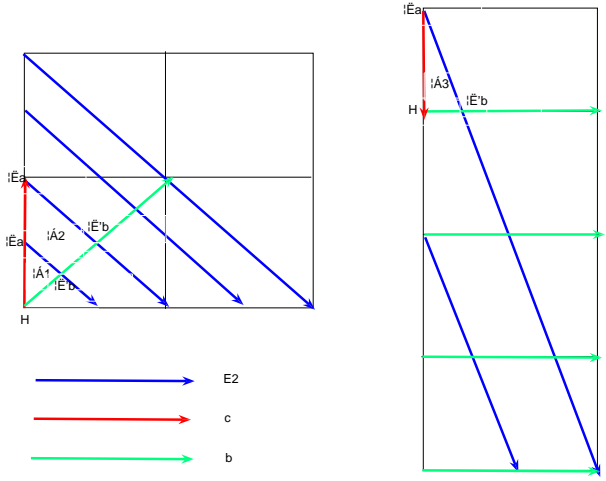}
\caption{\small The left and right diagrams correspond to
triangles on the first and second tori, respectively. In the first
torus, $E2$ and $c$ intersect twice,
$\mathcal{A}_{1},\mathcal{A}_{2}$ represent their areas. On
the second and third tori, they intersect only once, whose areas
are represented by $\mathcal{A}_{3}$ and $\mathcal{A}_{4}$.}
\label{triangle}
\end{figure}

The general strategy to compute charged matters coupling in $E2$
instanton background has been outlined in \cite{BCW}. In our case,
\begin{eqnarray}{\label{quartic}}
<(H_{u}H_{d})^{2}>_{E2-inst}&=&\int d^{4}x \sum_{conf}\prod_{a}
(\prod_{i=1}^{4}d\lambda_{ai})\times(\prod_{j=1}^{4}d\bar{\lambda}_{aj})e^{-S_{inst}}e^{Z'}\nonumber\\
&\times&<H_{u}>_{\lambda_{a1},\bar{\lambda}_{a1}}
<H_{d}>_{\lambda_{a2},\bar{\lambda}_{a2}}
<H_{u}>_{\lambda_{a3},\bar{\lambda}_{a3}}
<H_{d}>_{\lambda_{a4},\bar{\lambda}_{a4}}
\end{eqnarray}
which can be computed via conformal field theory techniques (see also \cite{Akerblom,Billo}).
To appreciate the structure of Eq. \eqref{quartic}, we take for example\footnote
{Other choices of $n^{1}_{N}$ and $n^{2}_{N}$ will yield more complex expressions,
but similar physics.}
$n^{1}_{N}=2,~ n^{2}_{N}=1$. They are shown by three simple
triangles in figure 1.
Non-perturbative terms in each torus are proportional to
\begin{eqnarray}
\left(H^{ij}_{u}\lambda_{a}^{i}\bar{\lambda}_{b}^{j}e^{-\mathcal{A}_{1}}+
H^{ij}_{d}\lambda_{a}^{i}\bar{\lambda}_{b}^{j}e^{-\mathcal{A}_{1}}\right)&+&
\left(H^{ij}_{u}\lambda_{a}^{i}\bar{\lambda}_{b}^{j}e^{-\mathcal{A}_{2}}+
H^{ij}_{d}\lambda_{a}^{i}\bar{\lambda}_{b}^{j}e^{-\mathcal{A}_{2}}\right),\nonumber\\
H^{ij}_{u}\lambda_{a}^{i}\bar{\lambda}_{b}^{j}e^{-\mathcal{A}_{3}}&+&
H^{ij}_{d}\lambda_{a}^{i}\bar{\lambda}_{b}^{j}e^{-\mathcal{A}_{3}},\nonumber\\
H^{ij}_{u}\lambda_{a}^{i}\bar{\lambda}_{b}^{j}e^{-\mathcal{A}_{4}}&+&
H^{ij}_{d}\lambda_{a}^{i}\bar{\lambda}_{b}^{j}e^{-\mathcal{A}_{4}},
\end{eqnarray}
respectively. $\mathcal{A}_{i}$ is the area in string units of the
triangle as shown in the figure 1. Note that
$\mathcal{A}_{2}=4\mathcal{A}_{1}$. The mixing terms between
$H_{u}H_{d}$s are highly suppressed due to simplicities of
triangle structure on the second and third tori. This leads to
the following term in the four-dimensional effective action,
\begin{eqnarray}
S_{nonpert}=\frac{A}{4!M}\varepsilon_{ijkl}\varepsilon_{mnpq}
H^{im}_{u}H^{jn}_{u}H^{kp}_{v}H^{lq}_{v},
\end{eqnarray}
where
\begin{eqnarray}
A=\frac{\pi^{3}}{4}(\Gamma_{1+\theta_{E2b},1-\theta_{E2c},1-\theta_{bc}})
\sum_{i,j=1}^{3}e^{-2(\mathcal{\tilde{A}}_{i}+\mathcal{\tilde{A}}_{j})}
\end{eqnarray}
and
\begin{eqnarray}{\label{M}}
M=g_{s}M_{s}\mathcal{V}_{E_{2}}e^{S_{inst}(E2_{N})}
\end{eqnarray}
where $\mathcal{V}_{E_{2}}=Vol(E2)/l_{s}^{3}$,
$\mathcal{\tilde{A}}_{i}=\mathcal{A}_{i},(i\neq 1,2)$ and
$\mathcal{\tilde{A}}_{1,2}=In(e^{\mathcal{A}_{1}}+e^{\mathcal{A}_{2}})$.
The rescaling for charged zero modes $\lambda\rightarrow \lambda
\sqrt{\frac{2\pi}{g_{s}}}$ and the $g_{s}$ factor independence for
each disc imply that each disc diagram carries an overall
normalization factor $2\pi/g_{s}$ \cite{CRW, Akerblom}. Thus, one
gets
\begin{eqnarray}{\label{mu}}
\mu \sim g_{s}^{-1} M_{s}e^{-S_{inst}(E2_{M})}
\end{eqnarray}
Eqs. (\ref{M}) and (\ref{mu}) determine the significance of
non-perturbative stringy effects on Higgs physics. With $M_{s}\sim
10^{18}$ GeV and $\mathcal{V}_{E_{2M}}\sim\mathcal{V}_{E_{2N}}\sim 10^{-30}$, one has
$\mu\sim 100$ GeV and $M\sim 1$ TeV.
The $\mathcal{V}_{E_{2M,N}}$ values will increase
as $n_{N}^{1}$ and $n_{N}^{1}$ decrease (the ratio of $Vol_{E2N}/Vol_{E2M}$ is smaller).
 Without any fine tunning,
these mass scales are exactly in the range desired by
phenomenlogy. We will see in the next section, in particular,
tree-level Higgs masses can be greatly uplifed.

There are other possible $E2_{N}$ instanton induced charged
matter couplings. Note that for $I_{Nh}\neq 0$ and
$I_{hc}=I_{hc^{*}}=6 $, which would generate terms
$<\phi_{h}\phi_{h}>_{E2_{N}}$, $<\psi_{h}\bar{\psi}_{h}>_{E2_{N}}$
to hidden matters with correct charged zero modes and other
measure assignments. The nonzero intersecting number $I_{MN}$
implies the possible existence of 1PI diagrams of multi-instantons.
These effects are of higher order and will not be included in the present analysis.

\section{The Higgs tree-level spectrum}
In MSSM-like models, Higgs physics provides a good window to test
new physics. In general, Higgs masses are sensitive to
supersymmetric breaking hidden sectors. This has recently been
revisited in the four-dimensional effective field theory formalism
\cite{DTS}. Earlier discussions on this topic were present in
\cite{BCEN}. It is shown that Higgs masses, especially the mass of
$h$ can be substantially uplifted by one type of quartic couplings
that were inherited from hidden sectors or extra
dimensions\footnote{These operators can be constructed in
five-dimensional $N=1$ supersymmetric theory, in which the fifth
dimension is compactified on the orbifold $S^{1}/Z_{2}$. The MSSM
is founded to be the four-dimensional effective field theory \cite{LZ}.}.

In models with two Higgs doublets $H_{u}=(H_{u}^{+},H_{u}^{0})$ and
$H_{d}=(H_{d}^{0},H_{d}^{-})$, there are 8 real Higgs scalars,
three are eaten by the massive $W$ bosons,
leaving two CP even $h$ and $H$, a CP odd $A_0$ and two charged $H^{\pm}$ particles.
The most general form of scalar superpotential that contains operators of effective dimension less than 5
is \cite{haber,martin}
\begin{eqnarray}\label{tree potential}
V&=&\tilde{m}^2_{H_{u}}H^{\dag}_{u}H_{u}+\tilde{m}^2_{H_{d}}H^{\dag}_{d}H_{d}-
(m^2_{ud}H_{u}H_{d}+h.c.)+\frac{\lambda_{1}}{2}(H^{\dag}_{u}H_{u})^2+
\frac{\lambda_{2}}{2}(H^{\dag}_{d}H_{d})^2\nonumber\\
&+&\lambda_{3}(H^{\dag}_{u}H_{u})(H^{\dag}_{d}H_{d})+\lambda_{4}(H^{\dag}_{u}H_{d})(H^{\dag}_{d}H_{u})\nonumber\\
&+& \left(\frac{\lambda_{5}}{2}(H_{u}H_{d})^2+(\lambda_{6}
H^{\dag}_{u}H_{u}+
\lambda_{7}H^{\dag}_{d}H_{d})H_{u}H_{d}+h.c.\right)
\end{eqnarray}
where the $\mu$ term and quartic terms come from hidden
sectors that break supersymmetry in the visible sector in general BMSSM models.
In our model they have extra contributions of non-perturbative origin. Instead of writing the masses as
functions of soft terms $\tilde{m}$, it is
more convenient to express them as three new parameters. Two of
them are the VEVs of $H_{u}^{0}$ and $H_{d}^{0}$, the third is
$m_{A^{0}}$. The dimensionless parameters are in our case,
\begin{eqnarray}
\lambda_1&=&\lambda_2=\frac{g^{'2}+g^2}{4},~~~~~~~~\lambda_3=\frac{g^{2}-g^{'2}}{4},\nonumber\\
\lambda_4&=&-g^{2}/2,~~~~~~\lambda_5=0,~~~~~~~\lambda_6=\lambda_7=2\epsilon
\end{eqnarray}
The extra new parameter
\begin{eqnarray}
\epsilon=\frac{A}{4!}\left(\frac{\mu}{M}\right)^2
\end{eqnarray}
is due to non-perturbative effects, which is in the range of $0.01\sim 0.1$ for typical values of $M$ and $\mu$.
The modifications on Higgs masses can be expressed as the the functions of
$v$ and expansion of $\epsilon$.
\begin{eqnarray}
\delta m^2_{H_{\pm}}&=&0 \\
\delta
m^{2}_{h}&=&2v^2sin(2\beta)\left(2\epsilon+\frac{(m^2_{A^{0}}+m^2_{Z})\epsilon}
{\sqrt{(m^{2}_{A^{0}}-m^{2}_{Z})^{2}+4m^{2}_{A^{0}}m^{2}_{Z}sin(2\beta)}}\right)+O(\epsilon^2)\\
\delta
m^{2}_{H}&=&2v^2sin(2\beta)\left(2\epsilon-\frac{(m^2_{A^{0}}+m^2_{Z})\epsilon}
{\sqrt{(m^{2}_{A^{0}}-m^{2}_{Z})^{2}+4m^{2}_{A^{0}}m^{2}_{Z}sin(2\beta)}}\right)+O(\epsilon^2)
\end{eqnarray}
Taking $\mu\sim 200$ GeV, $tan\beta=5$, the LEPII Higgs mass bound
$m_{h}\geq 114$ GeV can be accommodated with the $\delta m_{h}$ at
tree level when $M$ is below 20 TeV.
If $tan\beta$ decreases, one has to decrease $M$ also to uplift the $h$ mass substantially.
However, for moderate value of $\tan\beta$,
there will be a lower bound on $M$ from precision experiments\footnote{$M$ is bounded below by
electro-weak precise observables (EWPO). For example, one can
obtain a constraint on $M$ from the Fermi constant $G_{F}$, the
masses of $m_{W} $ and $ m_{Z}$ \cite{pomarol1},
\begin{eqnarray}\nonumber
\frac{G_{F}}{G_{F}^{SM}}=\left(1-(sin^{4}\beta+2sin^2\beta-1)\frac{\pi^{2}m_{W}^{(ph)2}}{3M^2}\right)
\end{eqnarray}
where ${\frac{G_{F}^{SM}}{G_{F}}}=1^{+0.0088}_{-0.0083}$ and
$m_{W}^{(ph)}=80.39\pm0.06$ GeV. For $tan\beta=5$, one needs
$M\geq 2.17$ TeV.}.
On the other hand, the following constraint relation between Higgs masses is
unchanged,
\begin{eqnarray}
m^2_{H_{\pm}}=m^{2}_{A^{0}}+m^{2}_{Z}
\end{eqnarray}
In addition to uplifting $m_{h}$, these operators
introduce new Higgs-Higgsino interactions,
which provide new channels in neutralino and chargino decays.
Pontentially, these phenomenological implications provides interesting tests of string theory.

\section{conclusions}
In this paper, we have discussed the $E2$ instanton-induced
superpotentials associated with Higgs physics in toroidal
orentifolds of type IIA theories. All the moduli in flux compactifications
are stabilized, which is very important to start
with. Explicitly, we present a $\mathcal{N}=1$ supersymmetric
model including two $E2$-branes. They induce the required $\mu$
term and extra quartic couplings. The later can be used to uplift the mass of the
lightest Higgs boson, as expected from general analysis of
four-dimensional effective field theory. In our case, they are generated
by non-perturbative stringy instanton effects, instead of other mechanisms in the hidden sector.

The wrapping numbers of this model are described by two real numbers
$(n_{E}^{1},n_{E}^{2})$, which preserve the same supersymmetry as those of
$D6$-branes. The structure of the induced quartic couplings can be calculated explicitly.
For illustration, we have calculated a very simple example, in which the numbers of
triangles are less than two on each torus. In this simple setting, we had the extra benefit that
the mixing terms are highly suppressed. With moderate and large $tan\beta$,
the mass of the lightest Higgs boson can be uplifted substantially to meet the LEPII bound.

Note that in all known intersecting brane models,
to generate non-perturbative neutrino masses seem to forbid a non-perturbative $\mu$ term at the same time,
and vice versa. Because the requirement of absence of zero modes in $E2-E2^{'}$
makes it very hard to satisfy all tadpole constraints and
supersymmetric conditions, as point out in \cite{CRW}. In our case, we havs succeed to
generated the $\mu$ term and extra quartic couplings, but not the desired neutrino masses.
Hopefully, this can be remedied in future, without sacrificing too much other attractive features in this class of models.

\section*{Acknowledgement}
This work is supported in part by the National Science Foundation
of China (10425525).

\end{document}